\documentclass[aps,prd,twocolumn,showkeys,showpacs,superscriptaddress,floatfix]{revtex4}

\usepackage{graphics,epsfig}
\usepackage{epstopdf}
\usepackage{subfigure}
\usepackage{palatino}
\usepackage{changes}
\usepackage[colorlinks=true,linkcolor=blue,urlcolor=blue,citecolor=blue]{hyperref}
\usepackage[toc,page]{appendix}
\usepackage[normalem]{ulem}
\usepackage{adjustbox}
\usepackage{latexsym}
\usepackage{amsmath}
\usepackage{amssymb}
\usepackage{amsfonts}
\usepackage{footnote}
\usepackage{times}
\usepackage{graphicx}
\usepackage{dcolumn}
\usepackage{amsmath}
\usepackage{array}
\usepackage{wrapfig}
\usepackage{multirow}
\usepackage{tabularx}
\usepackage{comment}
\usepackage{verbatim}

\def\ber{\begin{eqnarray}}
\def\eer{\end{eqnarray}}
\def\beq{\begin{equation}}
\def\eeq{\end{equation}}

\begin{document}

\title{Weak Gravitational Lensing in Ricci-Coupled Kalb–Ramond Bumblebee Gravity: Global Monopole and Axion-Plasmon Medium Effects }

\author{Ali {\"O}vg{\"u}n}
\email{ali.ovgun@emu.edu.tr}
\affiliation{Physics Department, Eastern Mediterranean University, Famagusta, 99628 North
Cyprus via Mersin 10, Turkiye.}

\date{\today}

\begin{abstract}
In this paper, we study the influence of the axion-plasmon medium, as proposed in [10.1103/PhysRevLett.120.181803]\cite{Tercas:2018gxv}, on the optical properties of black holes in a Lorentz-violating spacetime containing a global monopole. Our primary aim is to provide a test for detecting the effects of a fixed axion-plasmon background within the framework of Ricci-coupled Kalb–Ramond bumblebee gravity. By extending the conventional Einstein–bumblebee model through a nonminimal coupling between the Kalb–Ramond field and the Ricci tensor, we demonstrate that the combined presence of a global monopole and Lorentz-violating effects induces significant modifications to the classical Schwarzschild lensing signature. Employing the Gauss–Bonnet theorem within an optical geometry approach, we derive an analytical expression for the deflection angle that incorporates both linear and quadratic contributions from the Lorentz-violating parameter and the monopole charge. Furthermore, we investigate how the axion-plasmon coupling alters light propagation, affecting key observable gravitational deflection angle. Our results indicate that these optical characteristics are notably sensitive to the axion-plasmon parameters, thereby offering promising observational signatures for probing new physics beyond standard general relativity.

\end{abstract}



\maketitle

\section{Introduction}

Lorentz invariance is one of the cornerstones of modern physics, forming the bedrock of both the Standard Model and General Relativity. However, a growing body of theoretical work—motivated in part by string theory, quantum gravity, and noncommutative field theories—suggests that Lorentz symmetry may only be an approximate symmetry at low energies, with violations becoming manifest near fundamental scales. In this context, effective field theories such as the Standard-Model Extension (SME) have been developed to systematically parameterize possible departures from exact Lorentz invariance through small background fields that pervade spacetime \cite{Manton:2024hyc,Kostelecky:2003fs,Kostelecky:1988zi,Kostelecky:2000mm}.

A particularly intriguing candidate for introducing Lorentz violation is the Kalb–Ramond field, an antisymmetric rank-two tensor field originally proposed in the seminal work by Kalb and Ramond \cite{Manton:2024hyc,Lessa:2020imi}. Unlike the familiar electromagnetic potential, which is a one-form coupling naturally to point particles, the Kalb–Ramond field generalizes gauge interactions to extended objects such as strings by coupling to their two-dimensional world-sheets. In many string-inspired models, this field emerges as a massless excitation alongside the metric tensor and dilaton, and it may acquire a nonzero vacuum expectation value via spontaneous symmetry breaking. Such a mechanism—often studied within the so-called bumblebee models—naturally leads to Lorentz-violating effects in the gravitational sector, thereby altering both the structure of spacetime and the propagation of matter and radiation \cite{Paul:2022mup,Lessa:2019bgi,Duan:2023gng,Ortiqboev:2024mtk,Rahaman:2023swt,Yang:2023wtu,Belchior:2025xam,Baruah:2025ifh,Alimova:2024bjd,Filho:2024tgy,Junior:2024vdk,Filho:2023ycx,al-Badawi:2024pdx}.

The coupling between the KR field and the Ricci tensor introduces a mechanism through which Lorentz violation can influence the geometry of space-time, potentially giving rise to gravitational effects that deviate from those predicted by Einstein’s general relativity. This is particularly intriguing because Lorentz invariance—the principle that the laws of physics remain the same regardless of an observer’s velocity or orientation—is a cornerstone of modern physics.  On the other hand,  a minimal model for a global monopole is constructed using a triplet of scalar fields that initially respect a global $O(3)$ symmetry, which is spontaneously broken to a $U(1)$ symmetry. Notably, the gravitational field generated by a global monopole produces a spacetime with a solid angle deficit \cite{Barriola:1989hx,Rhie:1990kc}. By relaxing this symmetry and a global monopole which produces a spacetime with a solid angle deficit, the model opens the door to new physical phenomena, such as anisotropic gravitational interactions where the strength or behavior of gravity might vary with direction relative to the LV background field \cite{Colladay:1996iz,Bluhm:2004ep,Maluf:2015hda,Belchior:2023cbl,Lambiase:2023zeo,Ovgun:2018ran,Ovgun:2018xys,Pantig:2024ixc,AraujoFilho:2024iox,Panotopoulos:2024jtn,Lambiase:2024uzy,Gullu:2020qzu,Li:2020dln,Fathi:2025byw,Maluf:2020kgf,Guo:2023nkd,Kuang:2022xjp,Maluf:2022knd,Caloni:2022kwp,Delhom:2019wcm,Delhom:2020gfv,Filho:2022yrk,AraujoFilho:2024ykw,Altschul:2009ae,Assuncao:2019azw,Maluf:2018jwc,Liu:2024oas,Heidari:2024bvd}
. These effects could manifest in observable ways, such as subtle alterations in the dynamics of galaxy formation, the orbits of celestial bodies, or the propagation of gravitational waves across the universe.

The solar corona, the Sun’s outermost atmosphere, is a dynamic region of plasma and magnetic activity. Its study is critical for understanding how light bends as it passes near the Sun—a phenomenon known as the deflection angle, famously predicted by general relativity. This bending is influenced by the corona’s complex plasma environment, where structures like plasmoids can alter light propagation \cite{Turyshev:2018gjj,PhysRevLett.36.1475,An:2023wij}. Researching these effects is vital for refining our models of light deflection, which in turn enhances space weather predictions. Accurate forecasts protect satellite operations, power grids, and communication systems from solar events, making this work directly relevant to our technology-dependent society.

Our aim is to further explore these interactions and constrain the axion-plasmon coupling, potentially illuminating the effects on black hole lensing. Axions are extremely light, weakly interacting particles that not only offer potential solutions to the cosmological constant and the strong CP problem but also serve as promising dark matter candidates. Within the solar corona, axions could interact with the plasma, forming an axion plasma that modifies the region’s electromagnetic properties. Such interactions might subtly alter the deflection angle, offering a unique observational signature. Their energy density scales with the expansion of the universe, mirroring dark matter behavior, though their exact contribution remains uncertain \cite{Tercas:2018gxv}. In astrophysical settings, axion-photon conversion in magnetic fields (via the Primakoff mechanism) can influence stellar cooling, and their interactions with magnetized plasmas—particularly near compact stars and black holes—enhance observable optical phenomena \cite{Tye:2017upp,Svrcek:2006yi,Sikivie:2020zpn,DiLuzio:2020wdo,Odintsov:2019evb,Odintsov:2020iui,CAST:2017uph,Rogers:2015dla,Er:2017lue}
.  Investigating this possibility not only deepens our understanding of high-energy plasmas but also connects particle physics with astrophysics, potentially unveiling the nature of dark matter.

Investigating the weak deflection angle in this black hole background is motivated by several compelling reasons \cite{Virbhadra:1998dy,Virbhadra:1999nm,Virbhadra:2007kw,Virbhadra:2008ws,Virbhadra:2022iiy,Virbhadra:2002ju,Perlick:2003vg,Adler:2022qtb,Virbhadra:2022ybp,Perlick:2015vta,Atamurotov:2022knb,Javed:2019ynm,Javed:2022gtz,Javed:2022fsn,Crisnejo:2018uyn,Atamurotov:2015nra,Schee:2017hof,Turimov:2018ttf,Rayimbaev:2019nqg,Vagnozzi:2022moj,Bambi:2019tjh,Vagnozzi:2019apd,Khodadi:2020jij,Cvetic:2016bxi}
. First, gravitational lensing in the weak field regime serves as a sensitive probe of the underlying spacetime geometry, offering a potential observational window into modifications introduced by Lorentz-violating fields and global monopole effects. 

Unlike traditional Schwarzschild black holes, the presence of a nonminimal coupling between the Ricci tensor and the Kalb–Ramond field, together with the global monopole's solid angle deficit, alters the curvature structure in subtle ways that can significantly affect light trajectories. This modification is expected to yield distinct lensing signatures that can be used to discriminate between standard general relativity and extended gravity theories.

This work is organized as follows. In Section II, we briefly review the black hole solution in Ricci-coupled Kalb–Ramond bumblebee gravity with a global monopole. In Section III, we derive the weak deflection angle in vacuum using the Gauss–Bonnet theorem. In Section IV, we investigate the influence of an axion-plasmon medium on the weak deflection angle. Finally, in Section VI, we summarize our findings and present our conclusions.

\section{Brief Review of Black hole in Ricci-coupled Kalb-Ramond bumblebee
gravity with global monopole }

 In this section, we will review the black hole in ricci-coupled Kalb-Ramond bumblebee
gravity with global monopole 
 \cite{Belchior:2025xam}. By noting that $\kappa=8\pi G_N$, with $G_N$ denoting the Newtonian gravitational constant, $\Lambda$ representing the cosmological constant, and $\varepsilon$ being the coupling constant between the Ricci tensor and the Kalb–Ramond (KR) field $B_{\mu\nu}$, the action of the model is \cite{Lessa:2019bgi,Yang:2023wtu}

 \begin{widetext}
\begin{align}\label{actionKR}
    S=\int d^4x\sqrt{-g}\bigg[\frac{1}{2\kappa}\bigg(R-2\Lambda+\varepsilon\, B^{\mu\lambda}B^\nu\, _\lambda R_{\mu\nu}\bigg)-\frac{1}{12}H_{\lambda\mu\nu}H^{\lambda\mu\nu}-V(B_{\mu\nu}B^{\mu\nu}\pm b^2)+\mathcal{L}_m\bigg].
\end{align}
\end{widetext}
 In natural units, this constant has the mass dimension $[\varepsilon]=M^{-2}$. Furthermore, the antisymmetric field strength of $B_{\mu\nu}$ is defined by
\begin{equation} 
H_{\lambda\mu\nu} = \partial_\lambda B_{\mu\nu} + \partial_\mu B_{\nu\lambda} + \partial_\nu B_{\lambda\mu}.
\end{equation}
It is important to note that, as required in the bumblebee framework, the action explicitly breaks gauge symmetry due to both the nonminimal coupling and the smooth potential 
\begin{equation}
V\big(B_{\mu\nu}B^{\mu\nu}\pm b^2\big),
\end{equation}
which induces spontaneous Lorentz symmetry breaking. This mechanism results in a nonzero vacuum expectation value (VEV) for the KR field, namely, 
\begin{equation}
\langle B_{\mu\nu}\rangle = b_{\mu\nu}.
\end{equation}
Moreover, $T_{\mu\nu}^m$ denotes the energy–momentum tensor associated with conventional matter.

\begin{align}
  T_{\mu\nu}^m=-\frac{2}{\sqrt{-g}}\frac{\delta (\sqrt{-g}\mathcal{L}_m)}{\delta g^{\mu\nu}}.  
\end{align}
Then one can write the Einstein field equation as follows
\begin{align}\label{ER}
R_{\mu\nu}-\Lambda g_{\mu\nu}=T_{\mu\nu}-\frac{1}{2}g_{\mu\nu}T,    
\end{align}
where 
\begin{equation}
T_{\mu\nu} = \kappa\left(T_{\mu\nu}^m + T_{\mu\nu}^B\right) + T_{\mu\nu}^{\varepsilon}
\end{equation}
defines the total energy–momentum tensor and $T$ denotes its trace. Importantly, this tensor is conserved as a consequence of the Bianchi identities.

On the other hand, the Lagrangian density for the global monopole is \cite{Barriola:1989hx}
\begin{align}
\mathcal{L}_m=\frac{1}{2}\partial_\mu\psi^a\partial^\mu\psi^a - \frac{\chi}{4}(\psi^a\psi^a-\eta^2)^2. 
\end{align}
Above, the index $a$ labels the scalar fields $\psi^a$, with $a=1,2,3$, and the parameters $\chi$ and $\eta$ denote the self-coupling constant and the symmetry-breaking energy scale, respectively. To model the monopole configuration, we adopt the ansatz
\begin{equation}
\psi^a = \eta\,\frac{x^a}{r}, \quad \text{with} \quad x^a x^a = r^2,
\end{equation}
which is a valid approximation outside the monopole core. Since our focus is on exploring static and spherically symmetric spacetimes under the influence of a nonvanishing vacuum expectation value (VEV) of the Kalb–Ramond (KR) field, we consider the metric
\begin{align}\label{metric}
    ds^2 = -f(r)\,dt^2 + \frac{dr^2}{f(r)} + r^2\left(d\theta^2 + \sin^2\theta\,d\phi^2\right).
\end{align}

For a global monopole, the energy–momentum tensor associated with the scalar fields is given by \cite{Rhie:1990kc,Dadhich:1997mh}
\begin{align}
T^{m}_{\mu\nu} = \mathrm{diag}\Bigg( f(r)\,\frac{\eta^2}{r^2},\, \frac{1}{f(r)}\,\frac{\eta^2}{r^2},\, 0,\, 0 \Bigg).
\end{align}

Once the matter sector is specified, we proceed to configure the Lorentz-violating KR field. For this purpose, we assume a pseudo–electric configuration in which the field $B_{\mu\nu}$ is frozen to its VEV, and hence we write \cite{Lessa:2019bgi, Yang:2023wtu}
\begin{align}\label{KRconfig}
b_{\mu\nu} = b_{01} = -b_{10} = \frac{\vert b \vert}{\sqrt{2}}.
\end{align}
This configuration ensures a constant norm, namely, 
\begin{equation}
b_{\mu\nu}b^{\mu\nu} = \vert b \vert^2,
\end{equation}
and one may readily verify that the corresponding field strength
\begin{equation}
H_{\lambda\mu\nu} = \partial_\lambda B_{\mu\nu} + \partial_\mu B_{\nu\lambda} + \partial_\nu B_{\lambda\mu}
\end{equation}
vanishes identically, thereby satisfying the equations of motion for $B_{\mu\nu}$. For the initial analysis, we focus on solutions without a cosmological constant and adopt a quartic potential of the form 
\begin{equation}
V(X) = \lambda\, X^2,
\end{equation}
where $X = B_{\mu\nu}B^{\mu\nu} - b^2$, and $\lambda$ is the corresponding coupling constant. Under the configuration \eqref{KRconfig}, this potential satisfies $V=V'=0$.

Introducing the dimensionless Lorentz-violating (LV) parameter 
\begin{equation}
\gamma = \frac{\varepsilon\,\vert b \vert^2}{2},
\end{equation}
the solution for the metric function becomes \cite{Belchior:2025xam}
\begin{align}\label{sol1}
    f(r) = \frac{1-\kappa\eta^2}{1-\gamma} - \frac{2M}{r}.
\end{align}
Here, the parameter $\gamma$ quantifies the degree of Lorentz symmetry violation induced by the nonzero VEV of the KR field. In the limit $\gamma = \eta = 0$, the standard Schwarzschild solution is recovered.

Furthermore, the event horizon of the black hole is shifted due to the presence of the LV parameter, and its radius is determined by
\begin{align}
r_h = \frac{2M(1-\gamma)}{1-\kappa\eta^2}.
\end{align}

This analysis clearly illustrates that the incorporation of a global monopole and Lorentz-violating effects leads to nontrivial modifications of the standard black hole structure, affecting both the geometrical and thermodynamical properties of the spacetime.
\begin{table}[h!]
\centering
\caption{Constraints on the Lorentz-violating parameter \(\gamma\) from Solar System tests \cite{Yang:2023wtu}. }
\label{tab:constraints}
\begin{tabular}{|c|c|}
\hline
\textbf{Solar System Test} & \textbf{Constraint on \(\gamma\)} \\ \hline
Mercury precession         & \(-3.7 \times 10^{-12} \leq \gamma \leq 1.9 \times 10^{-11}\) \\ \hline
Light deflection           & \(-1.1 \times 10^{-10} \leq \gamma \leq 5.4 \times 10^{-10}\)  \\ \hline
Shapiro time delay         & \(-6.1 \times 10^{-13} \leq \gamma \leq 2.8 \times 10^{-14}\)  \\ \hline \label{table}
\end{tabular}
\end{table}

The parameter \(\gamma\) is a dimensionless quantity that characterizes the effects of Lorentz symmetry violation Table \ref{table}, resulting from the non-zero vacuum expectation value of the Kalb-Ramond (KR) field permeating spacetime. Solar System experiments, including measurements of the Shapiro time delay, light deflection, and Mercury's perihelion precession, constrain the parameter \(\gamma\) to the range \(-6.1 \times 10^{-13} < \gamma < 2.8 \times 10^{-14}\)~\cite{Yang:2023wtu}. 

\section{Weak Deflection Angle using Gauss-Bonnet Theorem}
\label{wdagbt}
For the equatorial plane $\theta=\pi/2$, the line element in Eq.~(\ref{sol1}) under null geodesics reduces to an optical metric of the form
\begin{equation}
    \label{om}
    \mathrm{d}t^2 = \frac{\mathrm{d}r^2}{f(r)^2} + \frac{r^2}{f(r)}\,\mathrm{d}\phi^2,
\end{equation}
with the determinant of the optical metric given by
\begin{equation}
g = \frac{r^2}{f(r)^3}.
\end{equation}

To compute the deflection angle via the optical geometry, we use the Gauss-Bonnet theorem (GBT) as formulated by Gibbons and Werner \cite{Gibbons:2008rj}, and then studied by many authors \cite{Ishihara:2016vdc,Jusufi:2017mav,Werner:2012rc,Ovgun:2018fnk,Okyay:2021nnh,Ono:2017pie,Islam:2020xmy,Kumar:2020hgm,Crisnejo:2018uyn,Ishihara:2016sfv,Jusufi:2017lsl,Ovgun:2019wej,Belhaj:2020rdb,Li:2020wvn,Pantig:2022gih,Jusufi:2017hed,Ovgun:2020gjz,Arakida:2017hrm,Cvetic:2016bxi,Atamurotov:2022knb,Atamurotov:2021cgh,Wang:2021irh,Ali:2024vxa}. In this framework, light rays are treated as spatial geodesics within the optical metric, and their deviation from straight-line motion results in a topological effect.

The Gaussian curvature $\mathcal{K}$, which is proportional to the Ricci scalar computed from the nonzero Christoffel symbols, is given by
\begin{equation}
\mathcal{K} = \frac{R}{2}.
\end{equation}

In order to evaluate the deflection angle using the optical Gaussian curvature, we select a non-singular domain $\mathcal{D}_{R}$ whose boundary is composed of the light ray trajectory $\gamma_{\tilde{g}}$ and an auxiliary curve $C_R$ at large radial distance, i.e., 
\begin{equation}
\partial\mathcal{D}_{R} = \gamma_{\tilde{g}} \cup C_R.
\end{equation}
Alternatively, one may choose a non-singular region outside the light ray path with Euler characteristic $\chi(\mathcal{D}_{R})=1$. For such a region, the Gauss-Bonnet theorem is expressed as \cite{Gibbons:2008rj}
\begin{equation}
\label{gbtwb}
\iint_{\mathcal{D}_{R}} \mathcal{K}\,\mathrm{d}S + \oint_{\partial \mathcal{D}_{R}} \kappa\,\mathrm{d}t + \sum_{i} \theta_i = 2\pi \chi(\mathcal{D}_{R}),
\end{equation}
where $\kappa$ denotes the geodesic curvature and $\theta_i$ are the exterior jump angles at the boundary.

In the limit as $R \to \infty$, the jump angles at the observer and source, $\theta_{\mathcal{O}}$ and $\theta_{\mathcal{S}}$, tend to $\pi/2$, so that 
\begin{equation}
\theta_{\mathcal{O}} + \theta_{\mathcal{S}} \to \pi.
\end{equation}
Defining $\ddot{\gamma}$ as the unit acceleration vector and assuming the unit-speed condition $\tilde{g}(\dot{\gamma},\dot{\gamma}) = 1$, the geodesic curvature is given by
\begin{equation}
\kappa = \tilde{g}\left(\nabla_{\dot{\gamma}} \dot{\gamma},\,\ddot{\gamma}\right).
\end{equation}
Thus, Eq.~(\ref{gbtwb}) becomes
\begin{equation}
\iint_{\mathcal{D}_{R}} \mathcal{K}\,\mathrm{d}S + \oint_{C_R} \kappa\,\mathrm{d}t \overset{R\rightarrow\infty}{=} \iint_{\mathcal{D}_\infty} \mathcal{K}\,\mathrm{d}S + \int_{0}^{\pi+\hat{\alpha}} \mathrm{d}\varphi = \pi,
\end{equation}
where $\hat{\alpha}$ represents the deflection angle.

Since the light ray $\gamma_{\tilde{g}}$ is a geodesic, one has $\kappa(\gamma_{\tilde{g}})=0$. By choosing the circular curve $C_R$, defined by $r(\varphi)=R=$ constant, the geodesic curvature simplifies to
\begin{equation}
\kappa(C_R)= \left\vert \nabla_{\dot{C}_R}\dot{C}_R \right\vert,
\end{equation}
where the radial component of the covariant derivative is evaluated as
\begin{equation}
\label{12}
\left( \nabla_{\dot{C}_R}\dot{C}_R \right)^r = \dot{C}_R^{\varphi}\,\partial_{\varphi}\dot{C}_R^{r} + \tilde{\Gamma}^{r}_{\varphi\varphi}\, \left( \dot{C}_R^{\varphi} \right)^2.
\end{equation}
In the large $R$ limit, the first term vanishes, and the second term, under the unit-speed condition, yields
\begin{equation}
\lim_{R\rightarrow\infty} \kappa(C_R) = \lim_{R\rightarrow\infty} \left\vert \nabla_{\dot{C}_R}\dot{C}_R \right\vert \to \frac{1}{R}.
\end{equation}
Moreover, for very large $R$, the optical line element approximates as
\begin{equation}
\lim_{R\rightarrow\infty} \mathrm{d}t \to R\,\mathrm{d}\varphi.
\end{equation}
Thus, one obtains
\begin{equation}
\kappa(C_R)\,\mathrm{d}t \to \mathrm{d}\varphi.
\end{equation}

Adopting the straight-line approximation for the light ray as $r = b/\sin\varphi$, where $b$ is the impact parameter, the deflection angle can be determined by the Gibbons-Werner method via the Gauss-Bonnet theorem:
\begin{equation}
\label{int0}
\hat{\alpha} = -\int_{0}^{\pi} \int_{r=b/\sin\varphi}^{\infty} \mathcal{K}\,\mathrm{d}S,
\end{equation}
with the surface element given by
\begin{equation}
\mathrm{d}S = \sqrt{g}\,\mathrm{d}r\,\mathrm{d}\varphi.
\end{equation}

The optical metric corresponding to the black hole solution Eq. \ref{sol1} leads to a Gaussian curvature computed from the non-zero Christoffel symbols, which is found to be

\begin{equation}
    \label{gauc}
    \mathcal{K} = \frac{3M^2}{r^4} + \frac{\displaystyle \frac{2M}{\gamma-1} - \frac{2\eta^2\kappa M}{\gamma-1}}{r^3}.
\end{equation}

This expression for $\mathcal{K}$ encapsulates the contributions from both the mass $M$ and the parameters characterizing the Lorentz violation and global monopole, thereby revealing how modifications in the gravitational sector affect the deflection of light in the weak field regime.

Neglecting higher-order contributions, the above equations simplify to the following asymptotic expression for the deflection angle:
\begin{eqnarray}
    \hat{\alpha} \approx \frac{3 \gamma \eta^2 \kappa M}{b\,(\gamma-1)} - \frac{2 \eta^2 \kappa M}{b\,(\gamma-1)} + \frac{6 \gamma M}{b\,(\gamma-1)} - \frac{4 M}{b\,(\gamma-1)},
    \label{eq:GBTa}
\end{eqnarray}
or equivalently,
\begin{eqnarray}
    \hat{\alpha} \approx &\frac{4M}{b} -\frac{\gamma^2 \eta^2 \kappa M}{b} - \frac{2\gamma^2 M}{b} - \frac{\gamma \eta^2 \kappa M}{b} - \frac{2\gamma M}{b} \notag \\ &+ \frac{2\eta^2 \kappa M}{b}.
    \label{eq:GBTa2}
\end{eqnarray}

\begin{figure}[htp]
   \centering
    \includegraphics[scale=0.6]{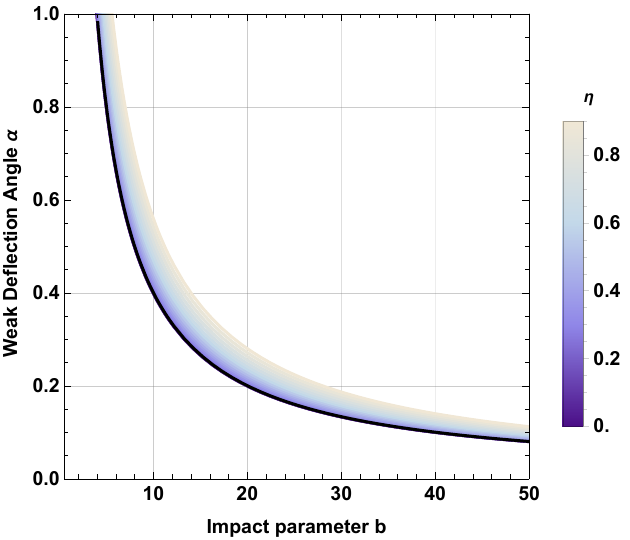}
    \caption{Figure shows the weak deflection angle $\hat{\alpha}$ versus impact parameter $b$ with $M=1$, $\gamma = 0.01$, $\kappa$ = 1,  for different values of $\eta$. The solid black line represents the Schwarzschild case.}
    \label{fig1}
\end{figure}

The Gauss-Bonnet theorem, when applied to a modified gravity model that incorporates Lorentz violation and a global monopole, provides an analytical expression for the weak deflection angle \(\hat{\alpha}\), shedding light on how gravitational, topological, and symmetry-breaking effects interact. This deflection angle scales inversely with the impact parameter \(b\) and depends on several key parameters: the black hole mass \(M\), the Lorentz-violating parameter \(\gamma\), and the global monopole parameters \(\eta\) and \(\kappa\). Specifically, the expression reveals that \(\gamma\) introduces linear and quadratic corrections that decrease \(\hat{\alpha}\), while the presence of the global monopole increases it shown in Figure \ref{fig1}, with mixed terms showing their combined influence. When \(\gamma = 0\) and \(\eta = 0\), the model reduces to the Schwarzschild case, confirming its consistency with standard general relativity. Graphical analysis further supports the \(1/b\) scaling and highlights how these parameters shape the deflection, suggesting that precise gravitational lensing observations could help measure them.

\section{Effect of the Axion-Plasmon Medium on the Deflection Angle}

The investigation of the axion-photon coupling is motivated by string theory considerations and the quest to elucidate dark matter properties. Incorporating axion-photon interactions into the electromagnetic framework opens up new theoretical avenues and novel phenomena. This generalization, inspired by works such as \cite{Mendonca:2019eke,Wilczek:1987mv,Atamurotov:2021cgh}, extends conventional electromagnetic theory by accounting for the influence of axions, which could have profound implications for both dark matter research and our broader understanding of fundamental interactions.

In our analysis, we adopt a generalized electromagnetic Lagrangian that includes axion-photon coupling:
\begin{equation}
\mathcal{L} = R - \frac{1}{4}F_{\mu\nu}F^{\mu\nu} - A_\mu J_e^\mu + \mathcal{L}_\varphi + \mathcal{L}_{\text{int}},
\end{equation}
where \(R\) is the Ricci scalar, \(F_{\mu\nu}\) the electromagnetic tensor, and \(J_e^\mu\) the electron four-current. The axion sector is described by the Lagrangian density
\begin{equation}
\mathcal{L}_\varphi = \nabla_\mu \varphi^* \nabla^\mu \varphi - m_\varphi^2 |\varphi|^2,
\end{equation}
while the interaction term
\begin{equation}
\mathcal{L}_{\text{int}} = -\frac{g}{4}\varepsilon^{\mu\nu\alpha\beta} F_{\alpha\beta} F_{\mu\nu}
\end{equation}
characterizes the photon-axion coupling, with \(g\) denoting the coupling constant.

The Hamiltonian governing the motion of a photon in the vicinity of a black hole immersed in an axion-plasmon medium is given by \cite{Synge:1960ueh}:
\begin{equation}
\mathcal{H}(x^\alpha, p_\alpha) = \frac{1}{2}\left[ g^{\alpha \beta} p_\alpha p_\beta - (n^2 - 1)(p_\beta u^\beta)^2 \right],
\label{generalHamiltonian}
\end{equation}
where \(x^\alpha\) are the spacetime coordinates, \(p_\alpha\) the photon's four-momentum, \(u^\beta\) its four-velocity, and \(n\) the refractive index (with \(n = \omega/k\), \(k\) being the wave number).

In the presence of an axion-plasmon medium, the refractive index takes the form \cite{Mendonca:2019eke}:
\begin{eqnarray}
n^2 &=& 1 - \frac{\omega_{\text{p}}^2}{\omega^2} - \frac{f_0}{\gamma_0}\frac{\omega_{\text{p}}^2}{(\omega - k u_0)^2} - \frac{\Omega^4}{\omega^2(\omega^2 - \omega_\varphi^2)} \nonumber \\
&& - \frac{f_0}{\gamma_0}\frac{\Omega^4}{(\omega - k u_0)^2(\omega^2 - \omega_\varphi^2)},
\label{eq:n1}
\end{eqnarray}
where the plasma frequency is defined as
\begin{equation}
\omega_{\text{p}}^2(x^\alpha) = \frac{4 \pi e^2 N(x^\alpha)}{m_e},
\end{equation}
with \(e\) and \(m_e\) being the electron charge and mass, respectively, and \(N\) the electron number density. The photon frequency is given by \(\omega^2 = (p_\beta u^\beta)^2\), and \(\omega_\varphi\) denotes the axion frequency. The axion-plasmon coupling parameter is
\begin{equation}
\Omega = \sqrt{gB_0 \omega_p},
\end{equation}
with \(B_0\) representing the homogeneous magnetic field in the \(z\)-direction. Here, \(f_0\) is the fraction of electrons in a beam propagating inside the plasma with velocity \(u_0\), and \(\gamma_0\) is the associated Lorentz factor. Given the uncertainties related to the electron beam scenario near the black hole, we simplify the analysis by setting \(f_0=0\). Equation (\ref{eq:n1}) then reduces to
\begin{eqnarray}
n^2(r) &=& 1 - \frac{\omega_{\text{p}}^2(r)}{\omega(r)^2} - \frac{\Omega^4}{\omega(r)^2\left[\omega(r)^2 - \omega_\varphi^2\right]} \nonumber \\
&=& 1 - \frac{\omega_{\text{p}}^2(r)}{\omega(r)^2} \left( 1 + \frac{g^2B_0^2}{\omega(r)^2 - \omega_\varphi^2} \right),
\label{eq:n2}
\end{eqnarray}
with the radial dependence of the photon frequency given by
\begin{equation}
\omega(r) = \frac{\omega_0}{\sqrt{A(r)}}, \qquad \omega_0 = \text{const}.
\end{equation}
Experimental investigations of axion-plasmon conversion typically impose the condition \(\omega_{\text{p}}^2 \gg \Omega^2\) or equivalently \(\omega_{\text{p}} \gg gB_0\) \cite{Mendonca:2019eke}.

A particularly straightforward model is one in which the medium is solely composed of axion-plasmon components. In this scenario, the refractive index is approximated as \cite{Atamurotov:2021cgh}
\begin{equation}
n(r) \simeq \sqrt{1 - \frac{\omega_{\text{p}}^2}{\omega_0^2} A(r) \left(1 + \frac{\tilde{B}_0^2}{1 - \tilde{\omega}_\varphi^2}\right)},
\label{45}
\end{equation}
which then allows us to reformulate the optical metric for a black hole embedded in a plasma:
\begin{equation}
dt^2 = \left[1 - \frac{\omega_{\text{p}}^2}{\omega_0^2} A(r) \left(1 + \frac{\tilde{B}_0^2}{1 - \tilde{\omega}_\varphi^2}\right)\right] \left[\frac{dr^2}{A(r)^2} + \frac{r^2}{A(r)} d\phi^2\right].
\end{equation}

The Gaussian curvature \(\tilde{\mathcal{K}}\) associated with the optical metric in this medium is derived from the non-zero Christoffel symbols and, after a lengthy calculation, can be expressed as

\begin{widetext}
\begin{eqnarray}
\tilde{\mathcal{K}}&=&\frac{2 \left(\left(\left(B_{0}^{2}-\omega_{\phi}^{2}+1\right) \omega_{p}^{2}-\frac{2 \omega_{0}^{2} \omega_{\phi}^{2}}{3}+\frac{2 \omega_{0}^{2}}{3}\right) \ln \! \left(r \right)+\left(-\frac{5 B_{0}^{2}}{3}+\frac{5 \omega_{\phi}^{2}}{3}-\frac{5}{3}\right) \omega_{p}^{2}+\omega_{0}^{2} \omega_{\phi}^{2}-\omega_{0}^{2}\right) \zeta q^{\frac{3}{2}}}{r^{3} \omega_{0}^{2} \left(\omega_{\phi}^{2}-1\right)}
\end{eqnarray}
\end{widetext}
where \(\zeta\) and \(q\) are model-dependent parameters.

Then the deflection angle can be obtained as
\begin{widetext}
\begin{eqnarray}
\alpha=-\int^{\pi}_0\int^{\infty}_{\frac{b}{\sin\phi}}\tilde{\mathcal{K}}dS 
 \simeq -\frac{B^2 \gamma  \eta ^2 \kappa  M \omega _e^2 \omega _{\psi }^2}{2 b \omega _{\infty }^2}-\frac{B^2 \gamma  \eta ^2 \kappa  M \omega _e^2}{2 b \omega _{\infty }^2}+\frac{B^2 \gamma  M \omega _e^2 \omega _{\psi }^2}{b \omega _{\infty }^2}+\frac{B^2 \gamma  M \omega _e^2}{b \omega _{\infty }^2}-\frac{B^2 \eta ^2 \kappa  M \omega _e^2 \omega _{\psi }^2}{b \omega _{\infty }^2}-\frac{B^2 \eta ^2 \kappa  M \omega _e^2}{b \omega _{\infty }^2} \notag \\+\frac{2 B^2 M \omega _e^2 \omega _{\psi }^2}{b \omega _{\infty }^2}+\frac{2 B^2 M \omega _e^2}{b \omega _{\infty }^2}-\frac{2 \text{$\gamma $M}}{b}-\frac{\gamma  \eta ^2 \kappa  M \omega _e^2}{2 b \omega _{\infty }^2}+\frac{\gamma  M \omega _e^2}{b \omega _{\infty }^2}-\frac{\eta ^2 \kappa  M \omega _e^2}{b \omega _{\infty }^2}+\frac{2 M \omega _e^2}{b \omega _{\infty }^2}-\frac{\gamma  \eta ^2 \kappa  M}{b}+\frac{2 \eta ^2 \kappa  M}{b}+\frac{4 M}{b}+\mathcal{O}(M^2,\gamma^2).~\label{deflangp}
\end{eqnarray}
\end{widetext}

\begin{figure}[h]
   \centering
    \includegraphics[scale=0.8]{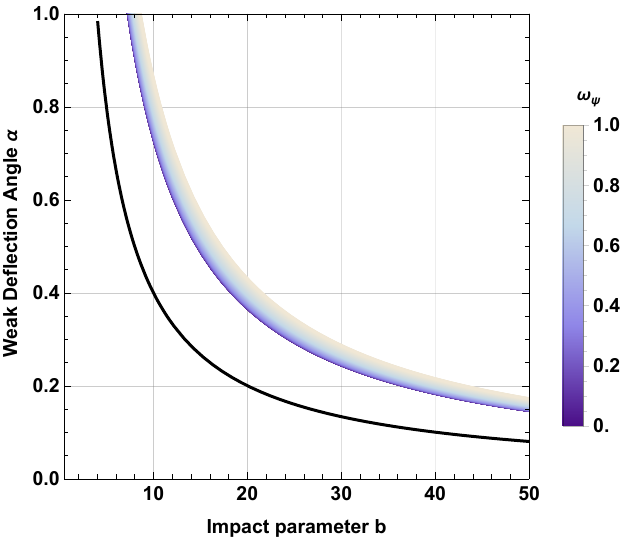}
    \caption{Figure shows $\alpha$ versus $b$ with $M=1$, $\gamma = 0.01$, $\kappa$ = 1, $z=0.9$, $B=0.9$ and $\eta=0.5$ for different values of $\omega_{\psi}$. The solid black line represents the Schwarzschild case.}
    \label{fig2a}
\end{figure}

\begin{figure}[h]
   \centering
    \includegraphics[scale=0.8]{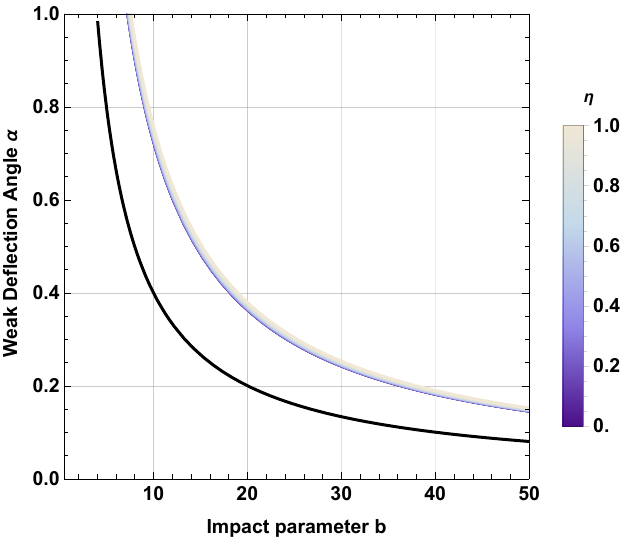}
    \caption{Figure shows the weak deflection angle $\hat{\alpha}$ versus impact parameter $b$ with $M=1$, $\gamma = 0.01$, $\kappa$ = 1, $\omega_{\psi}=0.1$ $\kappa$ = 1, $z=0.9$, and $B=0.9$   for different values of $\eta$. The solid black line represents the Schwarzschild case.}
    \label{fig2b}
\end{figure}

In this expression, the deflection angle \(\alpha\) is inversely proportional to the impact parameter \(b\), as expected in the weak-field limit, but it also contains several correction terms that depend on the magnetic field strength \(B\), the Lorentz-violating parameter \(\gamma\), the monopole parameter \(\eta\), the gravitational coupling \(\kappa\), and the characteristic frequencies \(\omega_e\), \(\omega_{\psi}\), and \(\omega_{\infty}\).

These additional terms reflect the influence of the axion-plasmon medium on the propagation of light in the vicinity of the black hole. In particular, the presence of both linear and quadratic contributions in \(\gamma\) and \(\eta\) indicates that the axion-photon interaction and the underlying plasma effects introduce nontrivial modifications to the standard gravitational lensing scenario. Such corrections may lead to observable deviations from the Schwarzschild prediction, thereby offering a potential window to probe new physics related to axion dynamics and plasma interactions in strong gravitational fields.

The behavior of the deflection angle as a function of the impact parameter is illustrated in Figures~\ref{fig2a} and \ref{fig2b} . For instance, of Fig.~\ref{fig2a} shows the variation of \(\alpha\) with \(b\) for a increasing value of \(\omega_{\psi}\), while the Fig.~\ref{fig2b} depicts how changes in the parameter \(\eta\) affect the deflection angle. Figures~\ref{fig2a} and \ref{fig2b}   further reinforces the sensitivity of the deflection angle to the combined effects of the axion-plasmon coupling and the Lorentz-violating parameters. Overall, these plots confirm the anticipated \(1/b\) decay in the asymptotic regime and demonstrate that even subtle modifications in the medium’s properties can lead to measurable changes in the lensing signature.

In summary, the result in Eq.~(\ref{deflangp}) and its corresponding plots underscore the significant impact of the axion-plasmon medium on the bending of light. Future high-precision gravitational lensing observations could leverage these deviations as a diagnostic tool for detecting axionic and plasma effects, thereby providing deeper insights into the interplay between modified electromagnetic interactions and gravitational phenomena.

\section{Conclusion}
\label{conc}

In this paper, we have investigated the weak deflection angle of a black hole within the framework of Ricci-coupled Kalb–Ramond bumblebee gravity, with a particular focus on the combined effects of a global monopole and Lorentz symmetry violation. By extending the conventional Einstein–bumblebee model, we have incorporated a nonminimal coupling between the Kalb–Ramond field and the Ricci tensor, leading to a modified gravitational action that naturally gives rise to Lorentz-violating effects. The presence of a global monopole, characterized by a triplet of scalar fields undergoing spontaneous symmetry breaking, introduces a solid angle deficit in the spacetime geometry. This combined setup modifies the classical Schwarzschild solution, yielding novel corrections to the gravitational lensing signature.

Using the Gauss–Bonnet theorem in the context of optical geometry, we derived an analytical expression for the weak deflection angle, which explicitly exhibits corrections that depend on the Lorentz-violating parameter \(\gamma\) and the monopole parameter \(\eta\). Our result, featuring both linear and quadratic terms in these parameters, indicates that the bending of light is not only governed by the black hole mass but also by the underlying modifications to the spacetime due to Lorentz violation and the global monopole. The deflection angle exhibits the characteristic \(1/b\) decay in the asymptotic regime, yet with additional correction terms that could serve as potential observational signatures.

Furthermore, we extended our analysis to include the effect of an axion-plasmon medium on the deflection angle. By incorporating axion-photon interactions inspired by string theory and dark matter research, we generalized the electromagnetic framework to account for axionic corrections. The resulting modifications to the refractive index and the optical metric lead to further alterations in the deflection angle. Our numerical plots demonstrate that the interplay between the axion-plasmon coupling, the Lorentz-violating effects, and the global monopole significantly influences the bending of light, thereby providing a rich structure that could be explored in future observational studies.

Overall, the findings presented in this work open up promising avenues for testing deviations from standard general relativity. High-precision gravitational lensing observations, by detecting subtle departures from the Schwarzschild predictions, may offer indirect evidence of Lorentz symmetry breaking, the presence of global monopoles, and axionic effects. These results not only deepen our understanding of modified gravity theories but also provide potential constraints on new physics beyond the standard model of cosmology and particle physics.

\acknowledgments

A. {\"O}. would like to acknowledge the contribution of the COST Action CA21106 - COSMIC WISPers in the Dark Universe: Theory, astrophysics and experiments (CosmicWISPers), the COST Action CA22113 - Fundamental challenges in theoretical physics (THEORY-CHALLENGES) and CA 23130 - Bridging high and low energies in search of quantum gravity (BridgeQG). I also thank EMU, TUBITAK and SCOAP3 for their support.

\bibliography{ref}
\bibliographystyle{apsrev}

\end{document}